\documentclass[a4paper,11pt]{article}
\pdfoutput=1
\usepackage{pos}
\usepackage{hyperref}
\usepackage{aas_macros}
\usepackage{wrapfig}

\begin{document}
\title{TELAMON: Effelsberg Monitoring of AGN Jets with Very-High-Energy Astroparticle Emissions}
\ShortTitle{TELAMON: Effelsberg Monitoring of AGN Jets with VHE Astroparticle Emissions}
\author*[a]{M.~Kadler}
\author[b]{U.~Bach}
\author[c]{D.~Berge}
\author[a]{S.~Buson}
\author[a]{D.~Dorner}
\author[d]{P.G.~Edwards}
\author[a]{F.~Eppel}
\author[e]{M.~Giroletti}
\author[a,g]{A.~Gokus}
\author[f]{O.~Hervet}
\author[a]{J.~He{\ss}d{\"o}rfer}
\author[h]{S.~Koyama}
\author[b]{A.~Kraus}
\author[b]{T.P.~Krichbaum}
\author[i]{E.~Lindfors}
\author[a]{K.~Mannheim}
\author[j]{R.~de Menezes}
\author[k]{R.~Ojha}
\author[b]{G.F.~Paraschos}
\author[c]{E.~Pueschel}
\author[a]{F.~Rösch}
\author[b]{E.~Ros}
\author[a]{B.~Schleicher}
\author[c]{J.~Sinapius}
\author[l]{J.~Sitarek}
\author[g]{J.~Wilms}
\author[m]{M.~Zacharias}
\affiliation[a]{Institut für Theoretische Physik und Astrophysik, Universit\"at W\"urzburg, Emil-Fischer-Stra{\ss}e 31, 97074 W\"urzburg, Germany; $^b$\,Max-Planck-Institut für Radioastronomie, Auf dem Hügel 69, 53121, Bonn, Germany; $^c$\,DESY, 15738 Zeuthen, Germany; $^d$\,CSIRO Astronomy and Space Science, PO Box 76, Epping, NSW, 1710, Australia; $^e$\,INAF-Istituto di Radioastronomia, Bologna, Via Gobetti 101, 40129, Bologna, Italy; $^f$\,Santa Cruz Institute for Particle Physics and Department of Physics, UCSC, Santa Cruz, CA 95064, USA; $^g$\,Dr. Karl Remeis-Observatory and Erlangen Centre for Astroparticle Physics, Universit\"at Erlangen-N\"urnberg, Sternwartstr.~7, 96049 Bamberg, Germany; $^h$\,Institute of Astronomy and Astrophysics, Academia Sinica, 11F of Astronomy-Mathematics Building, AS/NTU No. 1, Sec. 4, Roosevelt Rd, Taipei 10617, Taiwan, R.O.C.; $^i$\,Finnish Centre for Astronomy with ESO, University of Turku, FI-20014 University of Turku, Finland; $^j$\,Universidade de S\~ao Paulo, Departamento de Astronomia, Rua do Mat\~ao, 1226, S\~ao Paulo, SP 05508-090, Brazil; $^k$\,NASA HQ, Washington, DC 20546, USA; $^l$\,Department of Astrophysics, Faculty of Physics and Applied Informatics, University of Łódź , ul. Pomorska 149/153, 90-236 Łódź, Poland; $^m$\,LUTH, Université de Paris, 92190 Meudon, France \& CSR, NWU, Potchefstroom, 2520, South Africa}
\emailAdd{matthias.kadler@astro.uni-wuerzburg.de}

\abstract{We introduce the TELAMON program, which is using the Effelsberg 100-m telescope to monitor the radio spectra of active galactic nuclei (AGN) under scrutiny in astroparticle physics, namely TeV blazars and candidate neutrino-associated AGN. Thanks to its large dish aperture and sensitive instrumentation, the Effelsberg telescope can yield  radio data superior over other programs in the low flux-density ($S\nu$) regime down to several 10\,mJy. This is a particular strength in the case of TeV-emitting blazars, which are often comparatively faint radio sources of the high-synchrotron peaked type. 

We perform high-cadence high-frequency observations every 2-4 weeks at multiple frequencies  up to $\nu=44$\,GHz. This setup is well suited to trace dynamical processes in the compact parsec-scale jets of blazars related to high-energy flares or neutrino detections. Our sample currently covers about 40 sources and puts its focus on AGN with very-high-energy astroparticle emission, i.e., TeV blazars and neutrino-associated AGN. Here, we introduce the TELAMON program characteristics and present first results obtained since fall 2020.
}

\FullConference{37$^{\rm{th}}$ International Cosmic Ray Conference (ICRC 2021)\\
		July 12th -- 23rd, 2021\\
		Online -- Berlin, Germany}
\maketitle

\section{Introduction -- Radio Spectral Monitoring of AGN in Astroparticle Physics}
\noindent
Blazars are active galactic nuclei (AGN) that emit violently variable broadband emission from radio to $\gamma$-ray energies. 
With decreasing luminosity, the peaks of their characteristic double-humped broadband spectra are shifted upwards and the high-energy emission reaches the very-high-energy (VHE) regime at TeV gamma rays.
High-peaked BL\,Lac objects (HBLs) are canonically defined as sources whose primary (synchrotron) emission hump peaks above $10^{15}$\,Hz \cite{Padovani1995}. In extreme blazars, the primary emission peak can reach up even higher by up to two orders of magnitude \cite{Ghisellini1999,Biteau2020}. 
{Blazars are of utmost interest for astroparticle physics as possibly dominant sources of ultrahigh-energy cosmic rays and neutrinos} e.g., \cite{Hillas1984,Mannheim1995}. In particular, HBLs and extreme blazars have been considered in several recent theoretical works as relevant neutrino sources \cite{TavecchioGhiselliniGuetta2014,Padovani2015,Giommi2020}. 
Their radio Doppler factors are surprisingly often found to differ drastically from the Doppler factors derived from high-energy observations, e.g., \cite{PinerEdwards2018}. To explain this so-called \textsl{Doppler crisis of TeV blazars}, models have been proposed involving multiple zones on parsec scales, which can be investigated with coordinated deep multiwavelength and long-term monitoring observations \cite[e.g.,][]{Hervet2019}. 

Recently, it has been shown that {AGN radio monitoring programs can also play a key role in understanding very-high-energy neutrino emissions}. A tentative picture is emerging in which enhanced very-high-energy neutrino emission might be characteristically associated with AGN in flaring states \cite{Plavin2020,Hovatta2021}. This general behaviour has already been seen before in case of the three individual neutrino-candidate blazars PKS\,1424$-$418 \cite{Kadler2016}, TXS\,0506+056 \cite{kun2019}, and PKS\,1502+106 (ATel\,12996). 
Further radio data are urgently needed to consolidate this emerging picture. 

Because of their high peak frequencies, HBL blazars are generally faint radio sources and difficult to observe, especially in single-dish monitoring programs. 
With its frequency agility and sensitivity, the 100-m Effelsberg telescope is able to yield superior data as compared to smaller dishes. 
In the TELAMON (Tev Effelsberg Long-term Agn MONitoring) program, we want to characterize the variability properties of AGN with very-high-energy astroparticle emission. 
\textsl{Lindfors et al. \cite{Lindfors2016}} have pioneered such a study for bright VHE-emitting BL\,Lac objects based on OVRO 15\,GHz data. They found that simple single-zone emission models cannot explain their variability patterns and conclude that continuous high-sensitivity and densely sampled radio light curves are needed to separate different jet-emission zones. Our goal is to extend such studies with TELAMON to a larger sample and to higher radio frequencies using more-sensitive radio data. 


\section{Observational Setup}
\noindent
In the first year of our program, we have observed primarily with the S14mm and S7mm receivers, which delivered simultaneous data at 19, 21, 23, 25, 36, 39, 41, and 44\,GHz. 
We have typically used 8 subscans per cross scan for S14mm and 16 subscans for S7mm. 
To optimize the cadence and weather-dependent detection-rates at these high frequencies, we are now also using the S20\,mm receiver (14 and 17\,GHz) for the fainter targets.
Our regular observations can be complemented via interleaved additional target-of-opportunity observations to increase the cadence and/or frequency coverage during periods of special interest such as planned multiwavelength campaigns or source flares.
Example preliminary results are shown in Fig.~\ref{fig:mrk421_mrk501}-- \ref{fig:0658_1259}.

\section{The Sample}
\noindent
We have compiled a unique sample of TeV-detected and neutrino-candidate AGN. We  exclude bright low-peaked blazars, which are well covered in
other monitoring programs. As a selection criterion, we include all sources whose low-state flux density falls below 500\,mJy.
Sources south of $+30^\circ$ are also observed by ATCA in coordination with the TANAMI program \cite{Stevens2012}.
This leads to a sample that is complete (down to 10--20\,mJy) for HBLs and that includes a sufficient number of representatives from other source classes for comparison studies (see Table~\ref{tab:sources}). 
Newly detected sources are dynamically included  in our program if they fulfill our sample criteria.

\footnotesize
\begin{table}[hbt]
\caption{\small The TELAMON Sample of TeV-emitting and neutrino-candidate AGN. }
\centering
\footnotesize
\begin{tabular}{@{}cccccc@{\,\,}c@{\,}}
\hline\hline
ID  & Alternative\ & Class$^\textrm{a}$ & Sub- & $S_{14mm}^\textrm{c}$ &  Redshift & Remark$^\textrm{d}$ \\
  (J2000)   &    Name        &        & sample$^\textrm{b}$ & [mJy] & \\
\hline
0035+5950 & 1ES\,0033+595 & HBL & \textsc{i} & 75 & 0.086 & T \\
0112+2244 & S2\,0109+22 & IBL & \textsc{ii} & 1100 & -- & A \\
0214+5144 & TXS\,0210+515 & HBL & \textsc{i} & 150 & 0.049 & \\
0221+3556 & S3\,0218+35 & FSRQ & \textsc{i} & 500 & 0.68466 & \\
0222+4302 & 3C\,66A & HBL & \textsc{ii} & 1000 & 0.34 & T\\
0232+2017 & 1ES\,0229+200 & HBL$^*$ & \textsc{i} & 40 & 0.1396 & A, G, T\\
0303$-$2408 & PKS 0301$-$243 & HBL & \textsc{i} & 200 & 0.2657 & A \\
0316+4119 & IC\,310 & RG/HBL & \textsc{i} & 150 & 0.0189  & T\\
0416+0105 & 1ES\,0414+09 & HBL$^*$ & \textsc{i} & 50 & 0.287 & A, T \\
0507+6737 & 1ES\,0502+675 & HBL & \textsc{i} & 50$^{\,(1)}$ & 0.341 & T \\
0509+0541 & TXS\,0506+056 & IBL/HBL & \textsc{ii} & 1750 & 0.3365 & A, G, M, $\nu$, T\\
0521+2121 & RGB\,J0521+212 & IBL & \textsc{ii} & 375 & -- & A, T\\
0650+2502  & 1ES\,0647+250 & HBL & \textsc{i} & 100 & -- & A \\
0658+0637 & NVSS\,J065844+063711 & HBL & \textsc{i} & 125 & -- & A, $\nu$ \\
0811+0237 & 1RXS\,J081201.8+023735 & HBL$^*$ & \textsc{i} & $50^{\,(1)}$ & 0.1721  & A\\
0913$-$2103 & MRC\,0910$-$208 & HBL$^*$ & \textsc{i} & $135^{\,(1)}$ & 0.198017  & A\\
0955+3551 & 3HSP\,J095507.9+355101 & HBL$^*$ & \textsc{i} & ~10 & 0.557 & $\nu$\\
1015+4926 & 1ES\,1011+496 & HBL & \textsc{i} & 225 & 0.212 & $\nu$, T\\
1058+2817 & GB6\,J1058+2817 & HBL & \textsc{i} & 100 & 0.4793 & A, T \\
1104+3811 & Mrk\,421 & HBL$^*$ & \textsc{ii} & 375 &  0.031  & G, $\nu$, T\\
1136+7009 & Mrk\,180 & HBL & \textsc{i} & 175 & 0.045278 & \\
1145+1936 & 3C\,264 & RG & \textsc{i} & 325 & 0.021718 & A, T \\
1217+3007 & ON\,325 & HBL & \textsc{ii} & 450 & 0.131 & A \\
1221+2813 & W\,Comae & IBL & \textsc{ii} & 475 & 0.102 & A, T\\
1221+3010 & 1ES\,1218+304 & HBL$^*$ & \textsc{i} & 68 & 0.184 & A, T \\
1230+2518 & ON\,246 & IBL & \textsc{ii} & 400 & 0.135 & A \\
1422+3223 & OQ\,334 & FSRQ & \textsc{ii} & 775 & 0.681 & \\
1427+2348 & OQ\,240 & HBL & \textsc{ii} & 400 & 0.647 & A, M, $\nu$, T\\
1428+4240 & 1ES\,1426+428 & HBL$^*$ & \textsc{i} & 30 & 0.129 & G, T\\
1443+2501 & PKS\,1441+25 & FSRQ & \textsc{i} & 150 & 0.94 & A \\
1518$-$2731 & TXS\,1515$-$273& HBL & \textsc{i} & 225 & 0.1281 &  A \\
1542+6129 & GB6\,J1542+6129 & IBL & \textsc{ii} & 115 & 0.507 &  M, $\nu$ \\
1555+1111 & PG\,1553+113 & HBL & \textsc{i} & 300 & 0.49 & A, $\nu$, T\\
1653+3945 & Mrk\,501 & HBL$^*$ & \textsc{ii} & 1000 & 0.034 & G, T\\
1728+5013 & I\,Zw\,187 & HBL$^*$ & \textsc{i} & 125 & 0.055 & G, T\\
1743+1935 & 1ES\,1741+196 & HBL$^*$ & \textsc{i} & 175 & 0.084 & A, G\\
1813+3144 & B2\,1811+31 & FSRQ & \textsc{i} & 100 & 0.117 & $\nu$ \\
1943+2118 & HESS\,J1943+213 & HBL$^*$ & \textsc{i} & $\sim 20^{\,(2)}$ & -- & A \\
1958$-$3011 & 1RXS\,J195815.6$-$301119 & HBL$^*$ & \textsc{i} & $100^{\,(1)}$ & 0.119329  & A\\
1959+6508 & 1ES\,1959+650 & HBL$^*$ & \textsc{i} & 225 & 0.048 & G, T\\
2018+3851 & TXS\,2016+386 & HBL & \textsc{i} & 400 & -- & \\
2158$-$3013 & PKS\,2155$-$304 & HBL & \textsc{i} & 325 & 0.116 & A, T\\
2243$+$2021 & RGB J2243+203 & HBL$^*$ & \textsc{i} & $115^{\,(1)}$ & 0.119329  & A\\
2347+5142 & 1ES\,2344+514 & HBL$^*$ & \textsc{i} & 150 & 0.044 & G, T\\
\hline
\multicolumn{7}{l}{\footnotesize 
$^\textrm{a}$ FSRQ: flat-spectrum radio quasar -- LBL: low-peaked BL\,Lac -- IBL: intermediate-peaked BL\,Lac -- HBL: high-
} \\
\multicolumn{7}{l}{\footnotesize 
 peaked BL\,Lac (extreme blazars are marked as HBL$^*$) -- RG: Radio galaxy; $^\textrm{b}$ \textsc{i)} Observations in the 20\,mm and 
} \\
\multicolumn{7}{l}{\footnotesize 14\,mm bands, \textsc{ii)} Observations in the 14\,mm and 7\,mm bands;
$^\textrm{c}$ Median flux densities from our first 9 months of 
} \\
\multicolumn{7}{l}{\footnotesize observations at 14\,mm wavelength, or estimated from 1) NED, 2) Gregory \& Condon, 1991; $^\textrm{d}$ A: ATCA monitoring;   
} \\
\multicolumn{7}{l}{\footnotesize T: Frequent TeV observations or monitoring
by FACT, H.E.S.S., MAGIC or VERITAS; G: GMVA observations in 
}\\
\multicolumn{7}{l}{\footnotesize 2020; M: in coordination with MOJAVE; $\nu$: Positionally   associated with a high-energy IceCube neutrino.
}\\
\multicolumn{7}{l}{\footnotesize  
}\\
\end{tabular}
\\
\label{tab:sources}
\end{table}

\normalsize

\section{High-frequency radio observations of TeV Blazars}
\begin{figure}
    \centering
    \includegraphics[width=0.49\columnwidth]{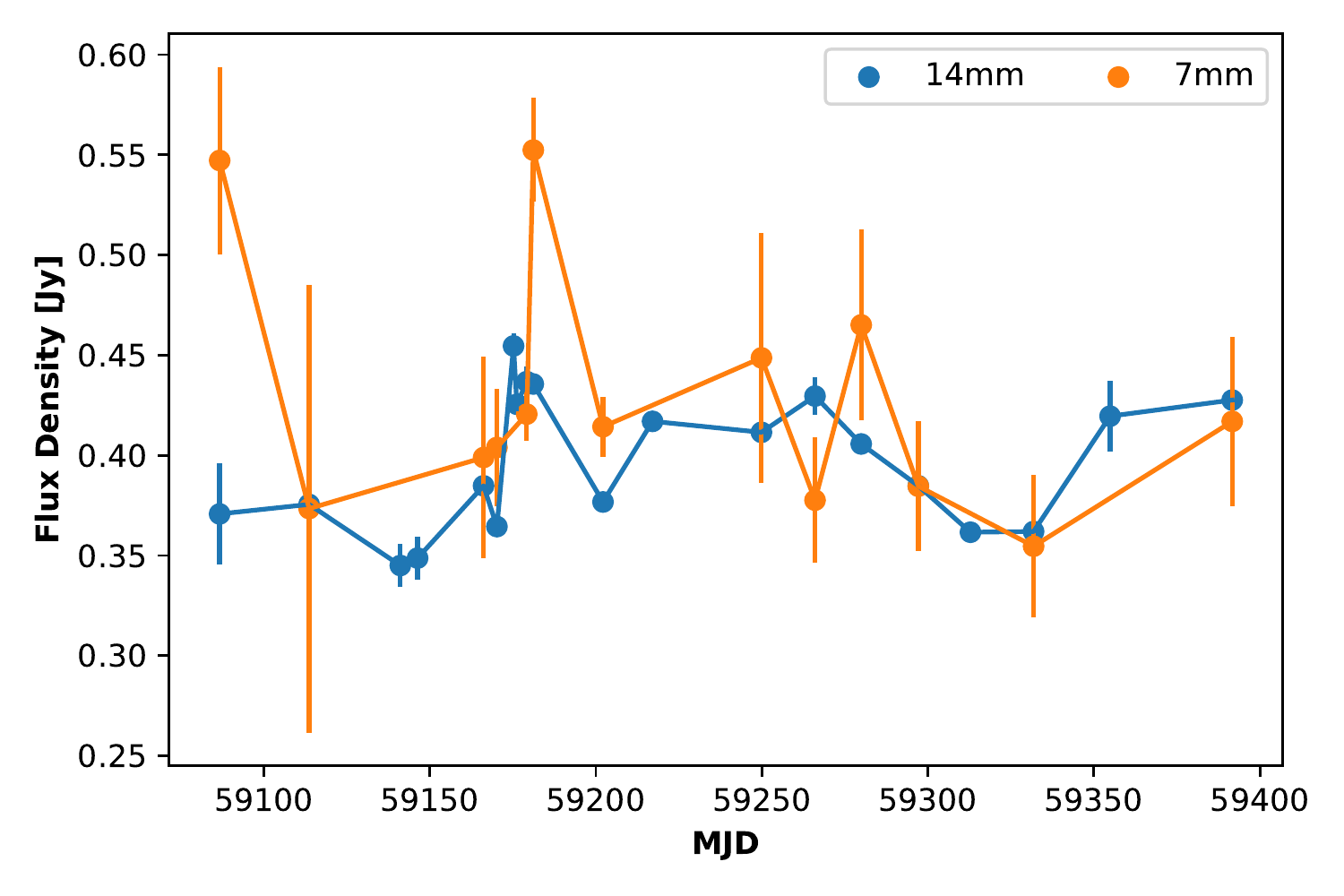}
    \includegraphics[width=0.49\columnwidth]{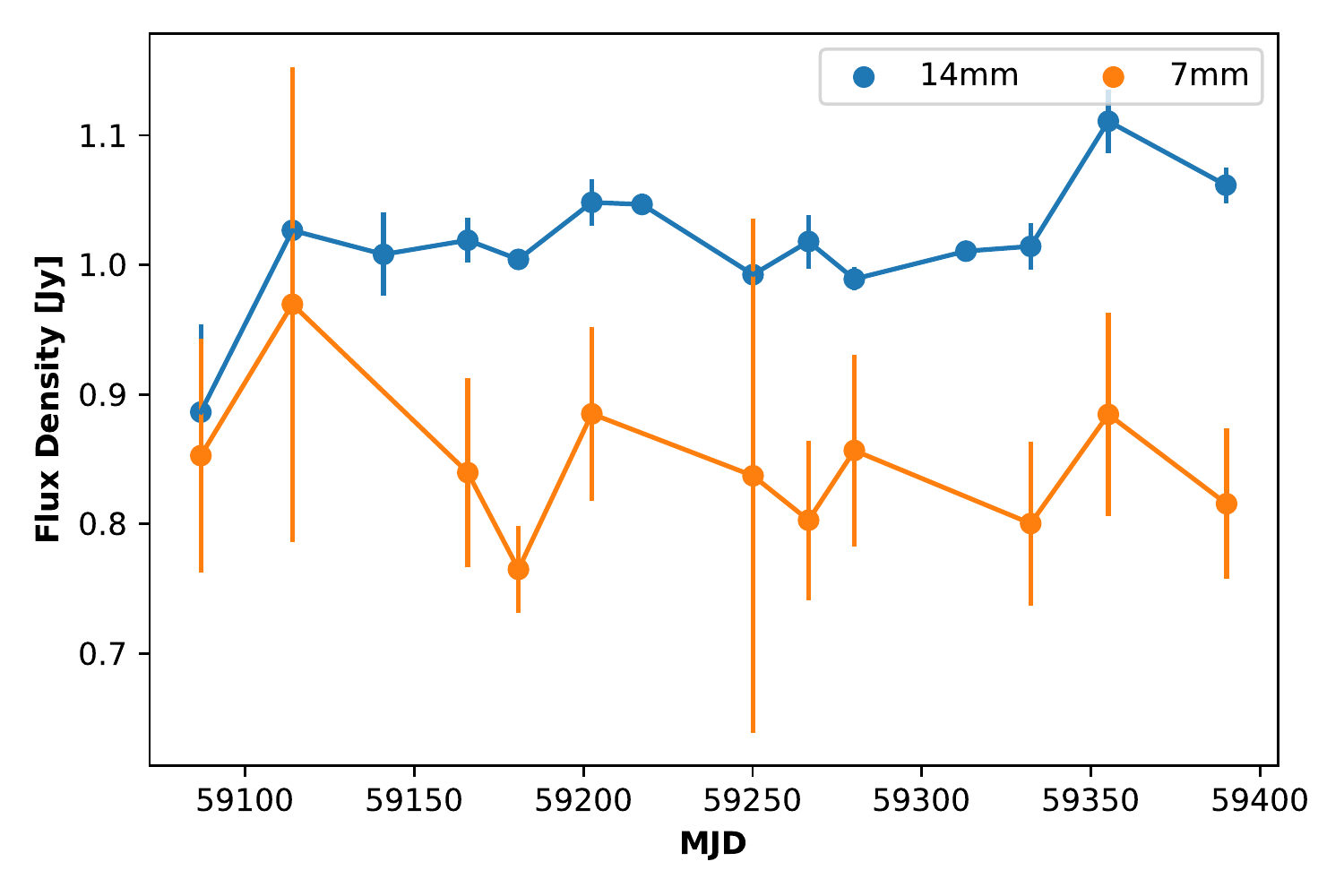}
    \caption{\small \sl TELAMON light curves (averaged over all subbands) of Mrk\,421 (left) and Mrk\,501 (right) between Sep 2020 and May 2021. }
    \label{fig:mrk421_mrk501}
\end{figure}

\begin{figure}
    \centering
    \includegraphics[height=4.6cm]{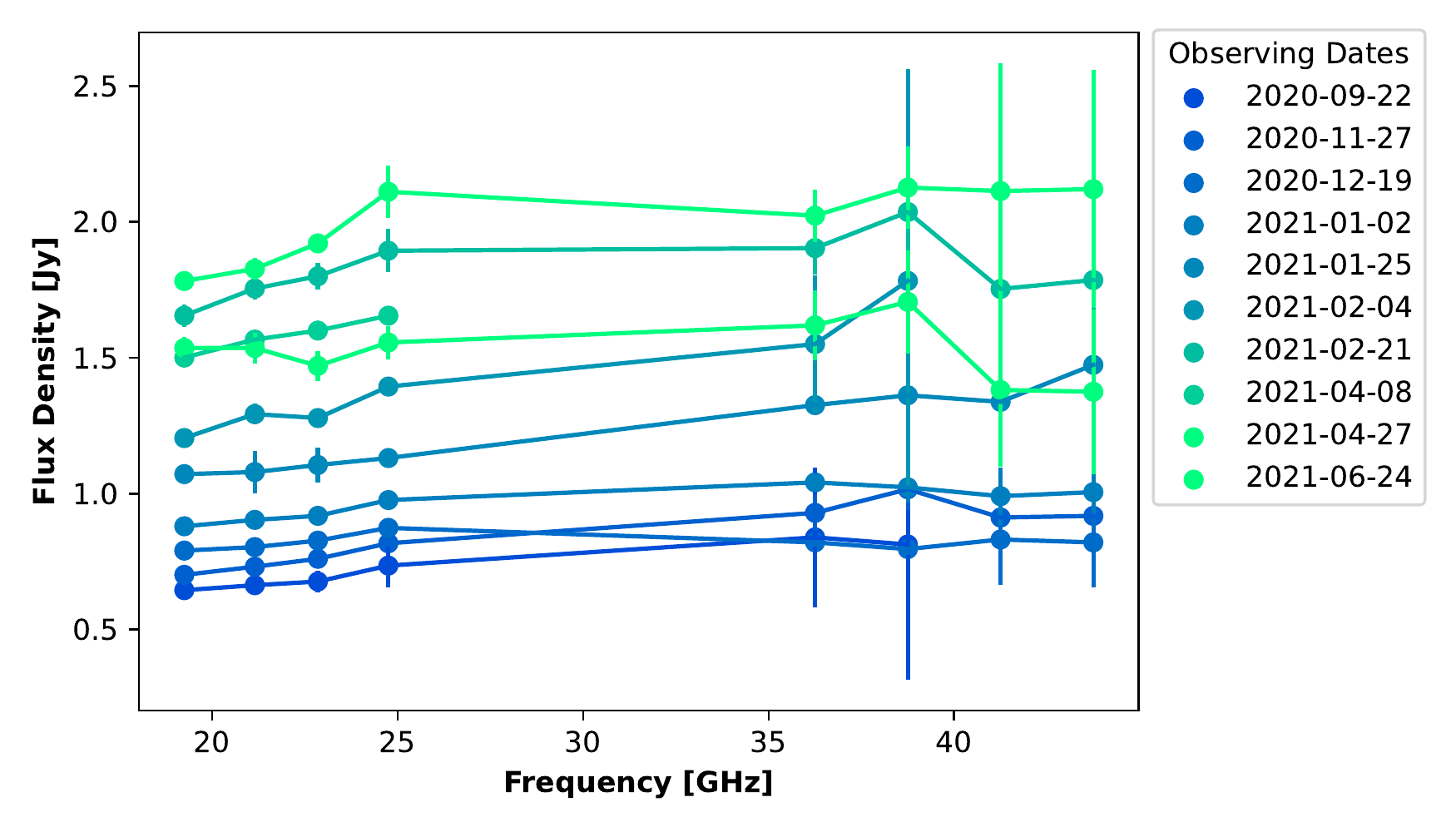}
    \includegraphics[height=4.6cm]{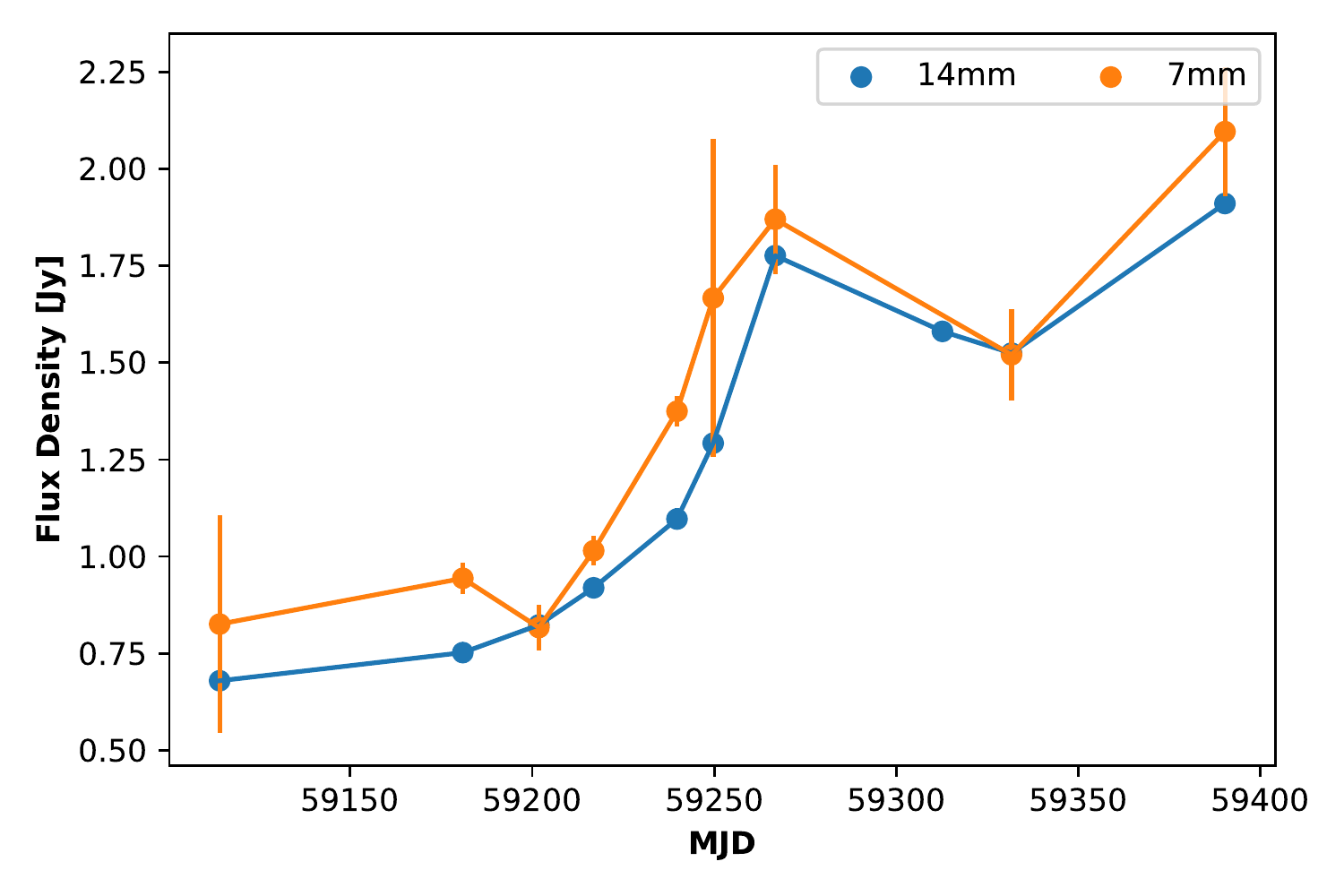}
    \caption{\small \sl Example spectra (left) and light curves (right, averaged over all subbands) for S2\,0109+22.}
    \label{fig:0112}
\end{figure}

\noindent
In Fig.~\ref{fig:mrk421_mrk501}, we show example light curves of the two well-known and fairly bright HBL blazars Mrk\,421 and Mrk\,501 throughout the first nine months of the TELAMON program. Both sources are frequent targets of TeV telescopes like FACT, MAGIC and VERITAS and show strong radio-jet variability in between gamma-ray observations. Continuous radio monitoring is especially important to put high-energy flaring results into context \cite[e.g.,][these proceedings]{gokus2021}.
In Fig.~\ref{fig:0112}, we show an example of a series of radio spectra for one of our program sources, S2\,0109+22.
This source shows flaring activity with a continuous increase in flux density over about 100 days both at 14\,mm and 7\,mm.

\section{High-frequency radio observations of neutrino-candidate AGN}
\noindent
In the first year of our program, TELAMON has contributed to the rapidly evolving field of neutrino astronomy by \textsc{i)} high-frequency spectral radio monitoring of the blazar TXS\,0506+056 that was found to show flaring activity in coincidence with the high-energy neutrino IceCube-170922A, and \textsc{ii)}  follow-up observations of three faint compact radio sources coincident with two newly detected IceCube neutrinos.

\textbf{\textsc{i)} TXS\,0506+056 --} On Sep 22, 2017, the IceCube telescope at the South Pole detected the 290\,TeV neutrino IceCube-170922A in spatial and temporal coincidence with flaring activity in the GeV gamma-ray band \cite{IceCube:2018dnn}. The chance coincidence was determined to less than $3 \sigma$ making this source the most compelling blazar association of any high-energy neutrino reported so far.

\begin{wrapfigure}{r}{0cm}
    \centering
    \includegraphics[width=0.5\columnwidth]{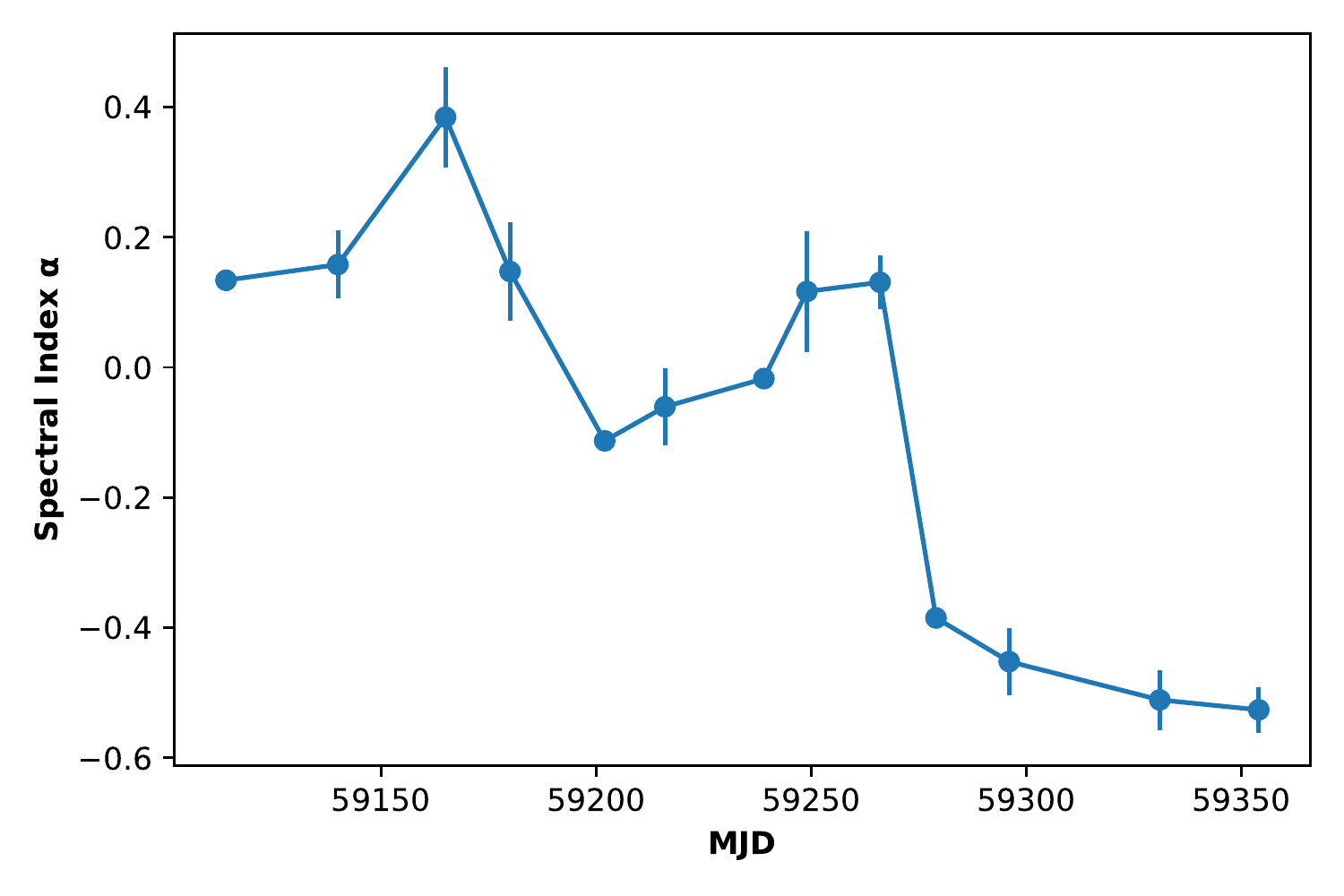}
    \caption{\small \sl 14--7\,mm spectral-index evolution of TXS\,0506+056 in early 2021. }
    \label{fig:txs0506}
\end{wrapfigure}

The long-term multiwavelength evolution of TXS\,0506+056 is shown in Satalecka et al. \cite[][these proceedings]{satalecka2021}. It shows that the radio emission of TXS\,0506+056 had already increased strongly months before the neutrino detection and has remained in a long-term outburst stage through 2020. This behaviour is similar to the blazar PKS\,1424$-$418 in association with the \textsl{BigBird} neutrino detected in 2012 \cite{Kadler2016}.
Recent radio-monitoring results of TELAMON (and also our associated ATCA program) show that this radio outburst seems to have ended in 2021, as indicated by a steeply decreasing trend at multiple frequencies over only a few months in early 2021 \cite[see][for the radio and multiwavelength light curves]{satalecka2021}.
In Fig.~\ref{fig:txs0506}, we show the radio spectral-index\footnote{The spectral index $\alpha$ is defined via $S_\nu \propto \nu^\alpha$.} evolution throughout this decrease. Through early 2021, the source still showed an inverted to flat spectrum. As of Mar 2021, the spectrum steepened significantly to values around $-0.4$ to $-0.5$. This is suggestive of a change into a jet state without fresh supply of high-frequency emitting plasma at the most-compact jet regions.

\begin{figure}
    \centering
    \includegraphics[height=4.6cm]{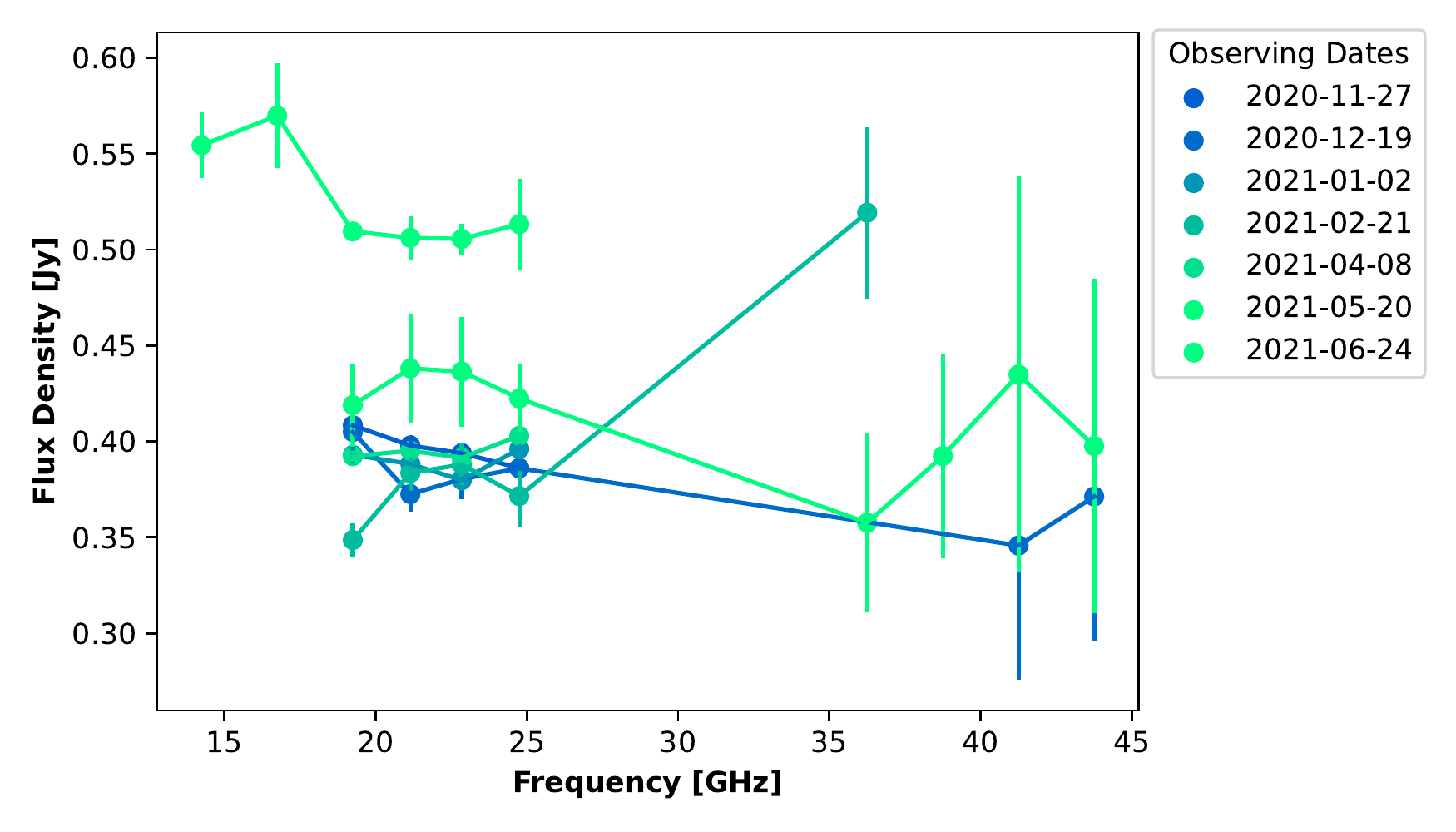}
    \includegraphics[height=4.6cm]{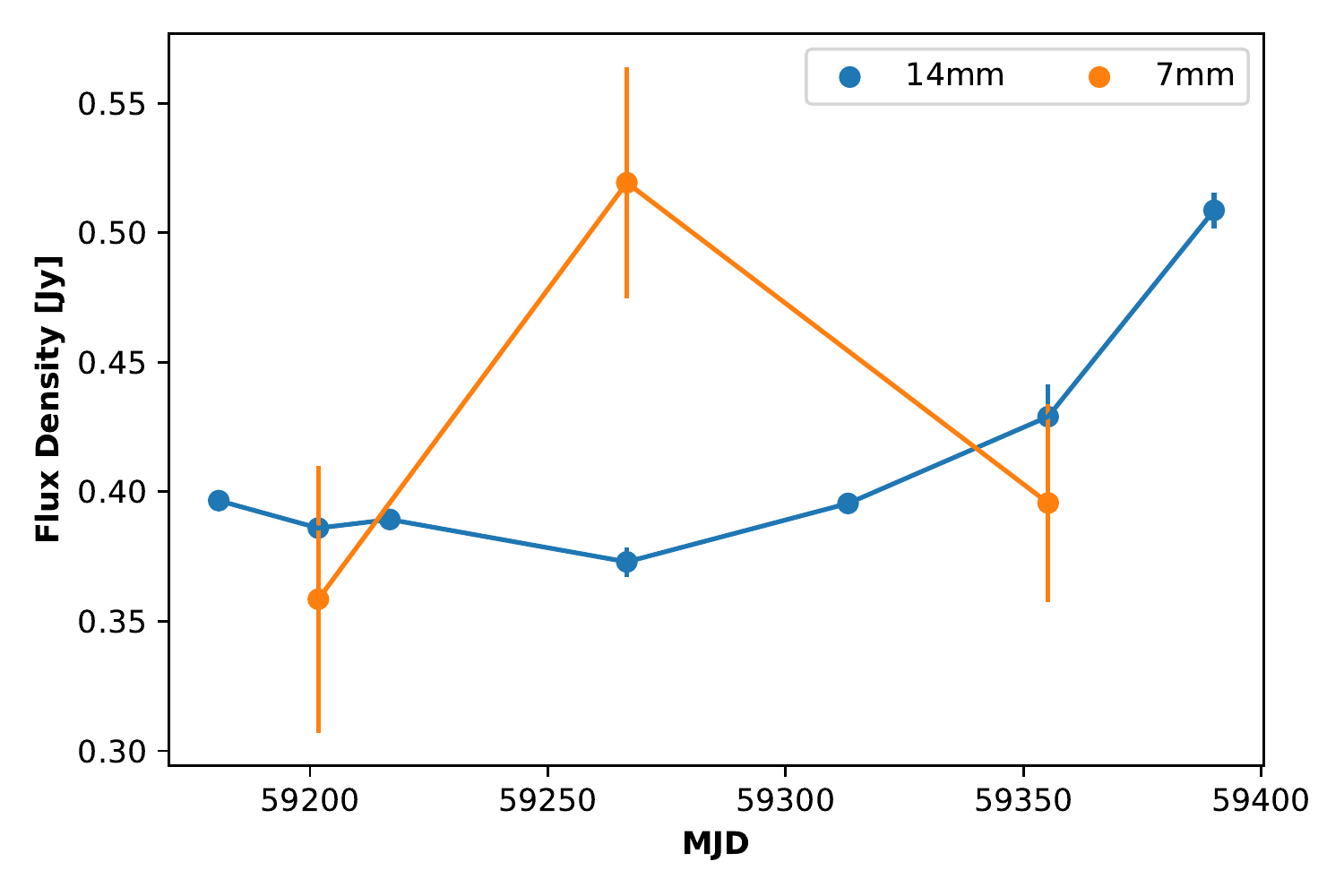}
    \caption{\small \sl Example spectra (left) and light curves (right, averaged over all subbands) for TXS\,2016+386.}
    \label{fig:2018}
\end{figure}

\textbf{\textsc{ii)} Radio follow-up observations of new blazar--neutrino candidate associations --} In mid/late 2020, the IceCube telescope reported the detection of the two bronze-alert events IceCube-201114A (GCN{\footnote{\url{https://gcn.gsfc.nasa.gov/gcn3_archive.html}}}\,28887) and IceCube-201120A (GCN\,28927, GCN\,28943), which TELAMON followed up in the radio band:

\textsl{TXS\,2016+386:} This is the radio-brightest AGN in the uncertainty region of IceCube-201120A. The 90\,\% neutrino 
localization is fairly large for this event. It covers about $85$ square degrees and is located near the 
Galactic plane (GCN\,28943). The positional association with TXS\,2016+386 is not highly significant as 13 other catalogued 4FGL gamma-ray sources are in the same field. We found radio flaring activity at 7\,mm (see Fig.~\ref{fig:2018}) followed by a pronounced increase at 14\,mm. Such observations, if supported by sufficient statistics in future events, can yield valuable additional information to judge the association significance of radio flaring and neutrino emission \cite[cf.][]{Hovatta2021}.

\begin{figure}
\centering
    \centering
    \includegraphics[width=0.49\columnwidth]{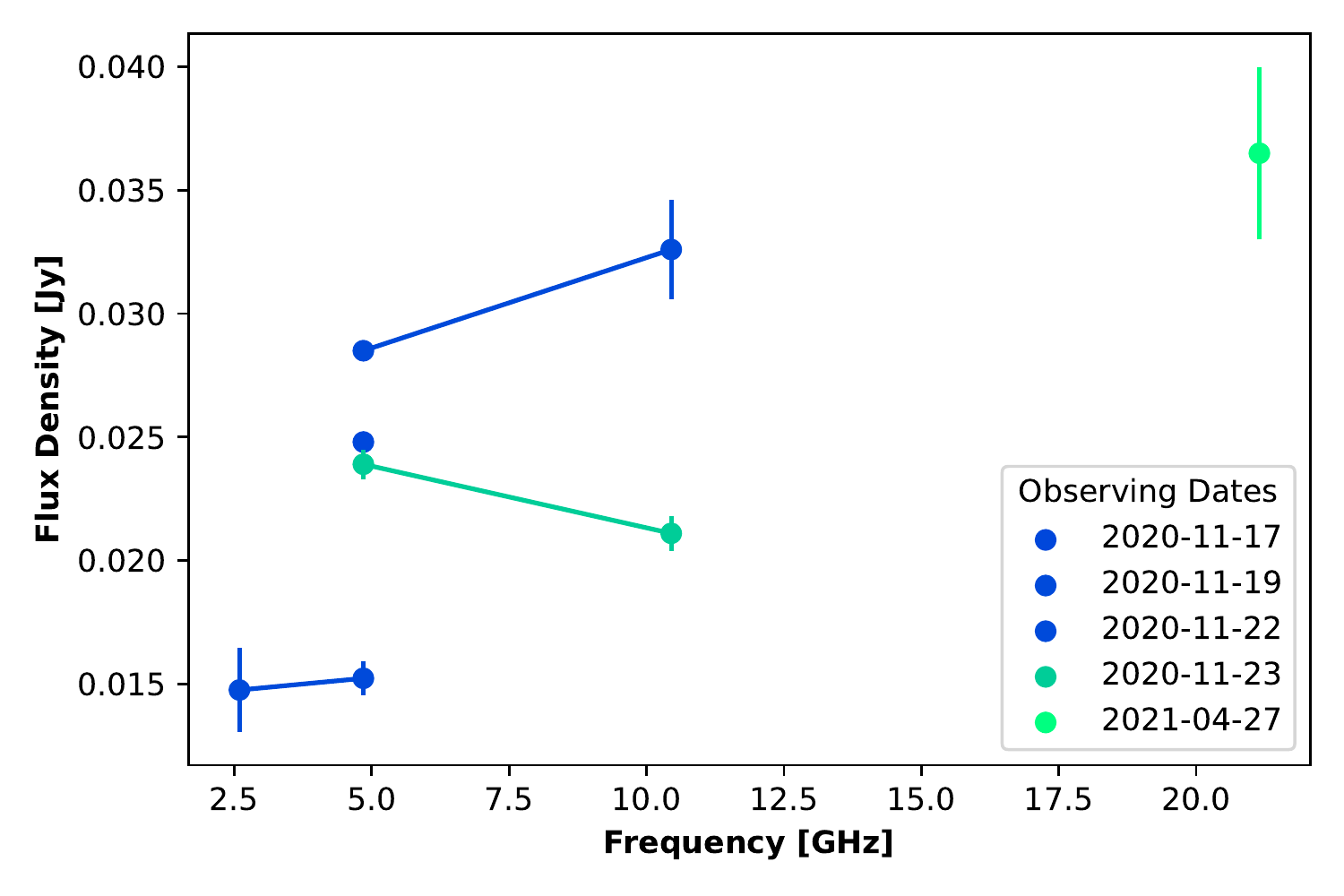}
    \includegraphics[width=0.49\columnwidth]{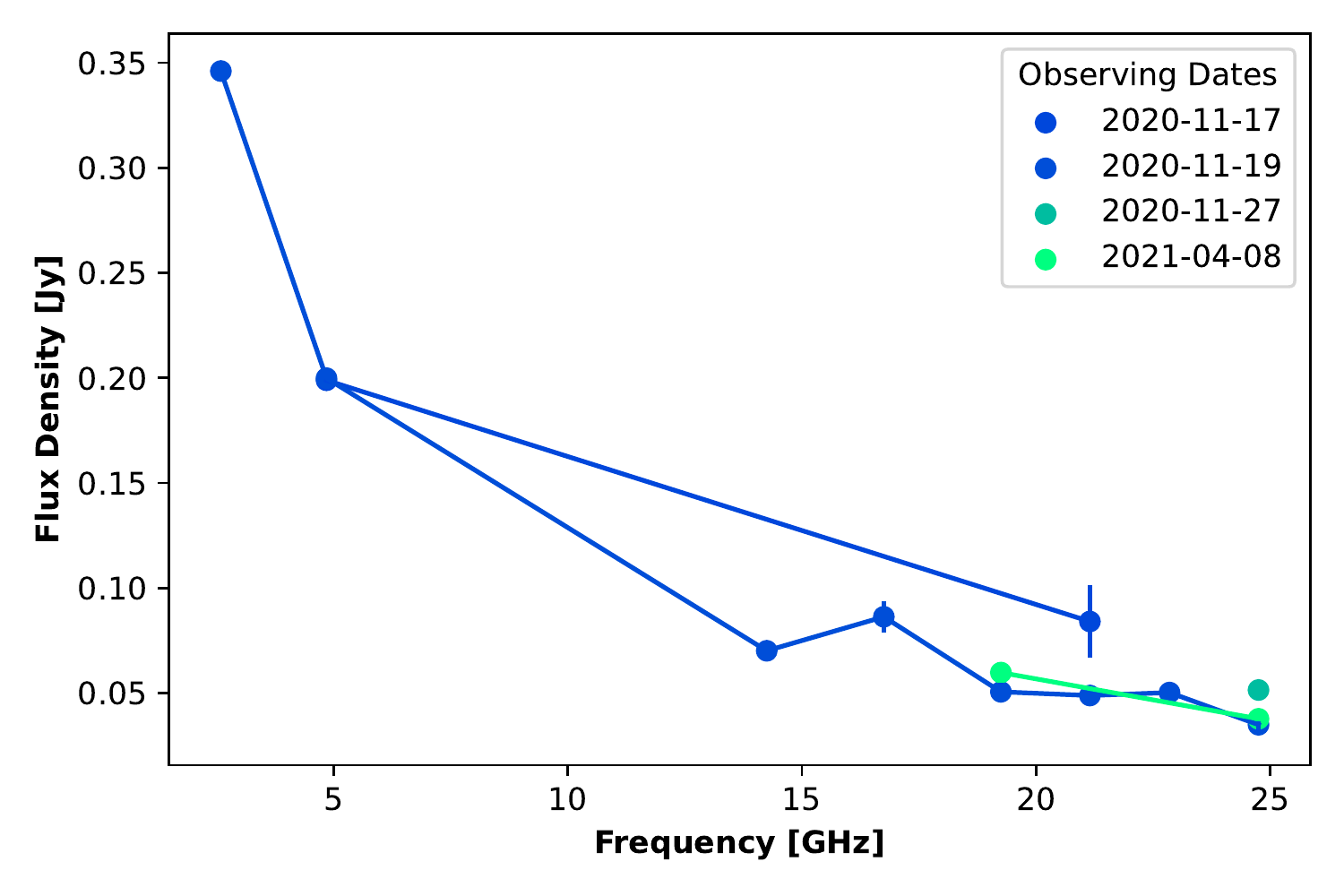}
    \caption{\small \sl Left: Radio spectral measurements of NVSS\,J065844+063711 performed three days after the detection of the IC\,201114A neutrino event.  Right: Radio spectrum of PKS\,1256+018.}
    \label{fig:0658_1259}
\end{figure}

\textsl{NVSS\,J065844+063711:} The association of this radio source with the neutrino IceCube-201114A is discussed in \cite[][these proceedings]{deMenezes2021}. This represents only the second candidate VHE object found within the 90\,\% confidence region of a well-reconstructed, high-energy IceCube event.
In the context of this multiwavelength effort, we could demonstrate the blazar-like flat radio spectrum of this formerly unclassified radio source (see Fig.~\ref{fig:0658_1259} and ATel\,14191).

\textsl{PKS\,1256+018:} We found that this source, which is located in the uncertainty region of the neutrino event IceCube-201115A, is an unlikely counterpart because it shows a steep radio spectrum (see ATel\,14191 and Fig.~\ref{fig:0658_1259}).

\section{Outlook}
\noindent
We are now at the dawn of a new era in high-energy astrophysics.
Current TeV gamma-ray instruments were able to detect  $\sim70$ AGN
and a few dozens of well localized high-energy neutrinos have been reported.
These numbers are small compared to the over 3000 AGN known in the GeV range \cite{Ajello2020} but will increase strongly with the advent of new major facilities.
The
Cherenkov Telescope Array (CTA, \cite{Acharya2013}) is planned to be completed in 2025 and
will observe between tens of GeV up to hundreds of TeV.
 In neutrino astronomy, IceCube continues to detect new events while, in the Northern Hemisphere, the new KM3NeT neutrino telescope is now coming online \cite{AdrianMartinez2016}.  The construction of much larger and more sensitive neutrino telescopes is planned: e.g., the IceCube-Gen2 will be about 10 times bigger than IceCube \cite{aartsen2021}.
Sensitive long-term monitoring radio programs like TELAMON with a focus on very-high-energy emitting AGN are clearly of high importance for the imminent CTA era as well as for the flourishing field of neutrino astronomy. 

\acknowledgments
\noindent
This research is based on observations with the 100-m telescope of the MPIfR (Max-Planck-Institut für Radioastronomie) at Effelsberg. 

\bibliographystyle{JHEP}
\bibliography{references}
\end{document}